\documentstyle [11pt]{article}
\begin{document}
%
%
\newcommand{\Abs}[1]{|#1|}
\newcommand{\Tr}[0]{\mbox{Tr}}
\newcommand{\EqRef}[1]{(\ref{eqn:#1})}
\newcommand{\FigRef}[1]{fig.~\ref{fig:#1}}
\newcommand{\Abstract}[1]{\small
   \begin{quote}
      \noindent
      {\bf Abstract - }{#1}
   \end{quote}
    }
\newcommand{\FigCap}[2]{
\ \\
   \noindent
   Figure~#1:~#2
\\
   }
\newcommand{\beq}{\begin{equation}}
\newcommand{\eeq}{\end{equation}}
%
%
%
%
\title{The Lyapunov exponent in the Sinai billiard 
in the small scatterer limit}
\author{
Per Dahlqvist \\
Mechanics Department \\
Royal Institute of Technology, S-100 44 Stockholm, Sweden\\[0.5cm]
}
\date{}
\maketitle

\ \\ \ \\

%
\Abstract{We show that Lyapunov exponent for the Sinai billiard
is $\lambda = -2\log(R)+C+O(R\log^2 R)$ with 
$C=1-4\log 2+27/(2\pi^2)\cdot \zeta(3)$ where $R$ is the radius of the
circular scatterer. We consider the disk-to-disk-map of the standard
configuration
where the disks is centered inside a unit square.
%
}

\ \\
\section{Introduction}

In this paper we will confirm, and make more precise 
a widely believed conjecture
concerning the Lyapunov exponent of the Sinai billiard 
in the small scatterer limit. We will find that the Lyapunov 
exponent for the disk to disk map is $\lambda = -2\log(R)+C+O(R\log^2 R)$
and derive an exact value of $C$ for a particular configuration of the
billiard, see below.
The Lyapunov number is known to be well defined for this system \cite{Sin}.
There has been a lot of good argument for the term
$-2\log R$ \cite{Oono,Chern,Dorf}
but the constant term has shown harder to approach.

The (largest) Lyapunov exponent is defined by
\begin{equation}
\lambda = \lim_{n \rightarrow \infty} \frac{1}{n} \log | \Lambda(x_0,n)|  \ \ ,
\end{equation}
defined for almost all initial points $x_0$.
$\Lambda(x_0,n)$ is the largest eigenvalue of the 
Jacobian of the n'th iterate of the map.

If the map is one-dimensional $x \mapsto f(x)$ then
\begin{equation}
\lambda = \lim_{n \rightarrow \infty} \frac{1}{n} \log | f^n(x_0)|
=\lim_{n \rightarrow \infty} \frac{1}{n} \sum_{i=0}^{n-1} \log |f'(x_i)|  \ \ ,
\end{equation}
which we, assuming ergodicity, can rewrite as
\begin{equation}
\lambda =  
\int  \log | f'(x) |\rho(x)dx  \ \ ,
\end{equation}
where $\rho(x)$ is the invariant density. The problem is that
$\rho(x)$ is usually unknown.

For an area-preserving map, as we will study, the problem is the reverse.
The invariant measure is known (it is uniform) but 
$\Lambda(x_0,n)$ does not have the simple multiplicative property.

Our strategy will be the following. 
We will first define a multiplicative weight
giving the correct $R\rightarrow 0$ behaviour of the Lyapunov exponent. Then
we will use the uniform measure to compute this limit. 
Central to the development is the distribution of recurrence times and
the majority of the paper will be spent on its derivation.

We will consider the {\em standard} configuration of the Sinai billiard
(although some of our considerations apply to slightly more general cases),
a unit square with a circular disk with radius $R$ centered on its midpoint.
We will study the disk to disk map, the Lyapunov exponent 
for the flow is available via
Abramov's formula \cite{Abr}.

\section{Calculations}

\subsection{Reducing the problem to one iterate of the map}

We are interested in computing the largest eigenvalue of the Jacobian of an
arbitrary long trajectory. For the time being,
we consider a rather general dispersive billiard with a circular disk with
radius $R$ inside a polygon and the associated disk to disk map.
The disk does not touch the polygon.
The Jacobian associated with one iterate of the map is, see e.g. \cite{SS}
\begin{equation}
J_i\equiv \left( \begin{array}{cc} 
\frac{\partial y'}{\partial y} & \frac{\partial y'}{\partial p_y}\\
\frac{\partial p'_y}{\partial y} &
\frac{\partial p'_y}{\partial p_y}\end{array}\right)= \left( 
\begin{array}{cc}
1+\frac{2l_i}{R\cos \alpha_i} & l_i\\
\frac{2}{R\cos \alpha_i} & 1 \end{array}
\right)  \ \ ,
\end{equation}
using a coordinate system following the trajectory - $y$ being the transverse
coordinate. 
$l_i$ is the traveling distance and $\alpha_i$ is the angle with respect
to the normal.
We have omitted an overall sign of $J$ which depend on the number
of times the particle hits the polygon during the disk to disk flight. 
For an area preserving mapping it is sufficient to know the trace 
of the Jacobian in order
to compute the eigenvalues so we
are interested in computing the trace of a product of matrices
$\prod_{i=1}^nJ_i$. 
We want to expand this trace in power series
in $R$ since we want to extract the $R\rightarrow 0$ limit of the Lyapunov
exponent. To this end it is convenient to split up $J_i$ into the following sum
\begin{equation}
J_i=\frac{2l_i}{R\cos\alpha_i } \left\{ \left( \begin{array}{cc}
1 & 0\\1/l_i & 0 \end{array} \right) + \frac{R\cos \alpha_i }{2}
  \left( \begin{array}{cc}
1/l_i & 1\\ 0 & 1/l_i \end{array} \right)  \right\}  \ \ ,
\end{equation}
or
\begin{equation}
J_i=\frac{2l_i}{R\cos\alpha_i } \left\{ A_i + \frac{R\cos \alpha_i }{2}
  B_i  \right\}  \ \ ,
\end{equation}
with
\begin{equation}
\begin{array}{cc}
A_i=\left( \begin{array}{cc}
1 & 0\\1/l_i & 0 \end{array} \right) & B_i=\left( \begin{array}{cc}
1/l_i & 1\\ 0 & 1/l_i \end{array} \right) \end{array}  \ \ .
\end{equation}
We write the trace as
\begin{equation}
\Tr \prod_{i=1}^{n}J_i=\Psi_n\cdot 
\prod_{i=1}^{n}\frac{2l_i}{R\cos (\alpha_i)}  \ \ , \label{eqn:Psidef}
\end{equation}
with
\begin{equation}
\Psi_n=\Tr\prod_{i=1}^n(A_i+\frac{R\cos\alpha_i }{2}B_i )  \ \ .
\end{equation}
The idea is now to show that $\Psi_n$ is bound close to unity when $R
\rightarrow 0$ and $n \rightarrow \infty$. 
First we use the fact that all matrix elements are strictly
positive and $\cos \alpha  <1$, so
\begin{equation}
1=Tr\prod_{i=1}^n A_i\leq \Psi_n\leq 
Tr \prod_{i=1}^n(A_i+\frac{R}{2}B_i ) \equiv
\Phi_n   \ \ .
\end{equation}
We expand $\Phi_n$ in powers of $R$
\begin{equation}
\Phi_n=a_{0,n}+a_{1,n}R \ldots a_{n,n}R^n  \ \ ,
\end{equation}
where for instance
\begin{equation}
a_{0,n}=1
\end{equation}
and
\begin{equation}
a_{1,n}= Tr(B_1 A_2 \ldots A_n)+Tr(A_1 B_2 \ldots A_n) \ldots +
Tr(A_1 \ldots A_{n-1}B_n)
\end{equation}
etc. The coefficient
$a_{m,n}$ is a sum over $(\begin{array}{c} n \\m \end{array} )$ combinations
of A's and B's.
A trace of such a combination in $a_{m,n}$
is a sum of terms like $1/(l_{i_1}l_{i_2}\ldots l_{i_m})$.
As all $l_i\geq l_{min}$ such a term is
$1/(l_{i_1}l_{i_2}\ldots l_{i_m})\leq 1/l_{min}^m$.
We can thus write
\begin{equation}
 \Phi_n \leq 1+\hat{a}_{1,n}(\frac{R}{l_{min}})+
\hat{a}_{m,n}(\frac{R}{l_{min}})^m+
\hat{a}_{n,n}(\frac{R}{l_{min}})^n  \ \ ,
\end{equation}
where
\begin{equation}
\hat{a}_{1,n}= Tr(\hat{B}\hat{A}^{n-1} )+Tr(\hat{A} \hat{B}  \hat{A}^{n-2}) 
\ldots +
Tr(\hat{A}^{n-1}\hat{B})  \ \ ,
\end{equation}
with
\begin{equation}
\begin{array}{cc}
\hat{A}=\left( \begin{array}{cc}
1 & 0\\1 & 0 \end{array} \right) & \hat{B}=\left( \begin{array}{cc}
1 & 1\\ 0 & 1 \end{array} \right) \end{array}  \ \ .
\end{equation}
Let us now take the trace of a particular combination of $\hat{A}$'s and $\hat{B}$'s.
This can always be organized  as
\begin{equation}
Tr(\hat{A}^{j_1}\hat{B}^{k_1}\hat{A}^{j_2}\hat{B}^{k_2}\ldots 
\hat{A}^{j_L}\hat{B}^{k_L})  \ \ ,
\end{equation}
\begin{equation}\begin{array}{ccc}
\sum_{\alpha=1}^{L} (j_\alpha+k_\alpha)=n & \mbox{  ,  }&
\sum_{\alpha=1}^{L} k_\alpha=m \end{array}  \ \ .
\end{equation}
The matrix $\hat{A}$ obey $\hat{A}^n=\hat{A}$ so we 
reorganize the trace above as
\begin{equation}
Tr(\hat{A} \hat{B}^{k_1}\hat{A} \hat{B}^{k_2}\ldots 
\hat{A}\hat{B}^{k_L}\hat{A})  \ \ .
\end{equation}
We also have $\hat{A}\hat{B}^k \hat{A}=(k+1)\hat{A}$ so the trace above is
bounded by
\begin{equation}
Tr(\hat{A} \hat{B}^{k_1}\hat{A} \hat{B}^{k_2}\ldots \hat{A}\hat{B}^{k_L}\hat{A})=
\prod_{\alpha=1}^{L} (k_\alpha +1)\leq 2^m  \ \ .
\end{equation}
So we finally have the following bound on
$\Phi_n$
\begin{equation}
1\leq \Psi_n \leq \Phi_n \leq
1+(\begin{array}{c}n\\1\end{array})(\frac{2R}{l_{min}})+
(\begin{array}{c}n\\m\end{array})(\frac{2R}{l_{min}})^m+\ldots
(\begin{array}{c}n\\n\end{array})(\frac{2R}{l_{min}})^n=(1+\frac{2R}{l_{min}})^n
  \ \ .  \label{eqn:Psibound}
\end{equation}

Let us now recall the definition of the Lyapunov exponent
\begin{equation}
\lambda =\lim_{n \rightarrow \infty}\frac{1}{n}\log |\Lambda(x_0,n)|  \ \ .
\end{equation}
The trace $\Tr J(x_0,n)$ for an area preserving mapping is
$\Tr J(x_0,n)=\Lambda(x_0,n)+1/\Lambda(x_0,n)$ so for large $n$ we have,
provided that the limit above exist,
$|\Lambda(x_0,n)|=|\Tr J(x_0,n)|+$ exponentially small terms.
Using eq. \EqRef{Psidef} we get
\begin{equation}
\lambda =\lim_{n \rightarrow \infty}\frac{1}{n}\log 
\prod_{i=1}^{n}(\frac{2l_i}{R\cos \alpha_i})+
\lim_{n \rightarrow \infty}\frac{1}{n}\log \Psi  \ \ ,
\end{equation}
and according to eq. \EqRef{Psibound} we get 
$0<\lim_{n \rightarrow \infty}\frac{1}{n}\log \Psi<2R/l_{min}$.
and
\begin{equation}
\lambda =\lim_{n \rightarrow \infty}\frac{1}{n} 
\sum_{i=1}^{n}\log(\frac{2l_i}{R\cos \alpha_i})+O(R)  \ \ ,
\end{equation}
saying that, to leading order, the Lyapunov 
exponent is just the arithmetic average
of a scalar. This enables us to rewrite it as phase space average
\begin{equation}
\lambda=\int \log |\frac{2l_s(x_s)}{R\cos \alpha(x_s)}|dx_s+O(R)  \ \ ,
\end{equation}
where $l_s(x_s)$ is the travel length to the next disk bounce and
$\alpha(x_s)$ the scattering angle as functions of the phase space point
$x_s$ on the disk. 
We choose as
coordinates for $x_s$
the scattering angle $\alpha$ and an angle
$\phi$ measured along the rim of the disk.
The invariant measure is then $dx_s=d\phi\;d (\sin \alpha )/4\pi$.

\subsection{Relating the Lyapunov to the distribution of recurrence times}

We can now divide the phase space integrals above into several terms
\begin{equation}
\lambda=\int \log(2/R) dx_s -\int \log\cos\alpha(x_s) dx_s+
\int \log l_s(x_s) dx_s +O(R)  \ \ .
\end{equation}
The first term is trivially $=\log(2/R)$. The second term is also easily
computed
\begin{equation}
\int \log\cos\alpha(x_s) dx_s= \frac{1}{2}\int_{-\pi/2}^{\pi/2}\log(\cos\alpha)
\;d(\sin\alpha )= \log(2)-1  \ \ .
\end{equation}
To evaluate the third term we insert the identity $1=\int \delta
(l-l_s(x_s))dl$ 
\begin{equation}
\int \log l_s(x_s) dx_s=\int d l \int dx_s \log (l_s(x_s))
\delta(l-l_s(x_s))=\int d l \log (l) p(l)  \ \ ,
\end{equation}
where
\begin{equation}
p(l)=\int dx_s \delta(l-l_s(x_s))   \label{eqn:pl}
\end{equation}
is the distribution of recurrence times.
We are going compute the small $R$ limit of this function for the standard
version of the Sinai billiard, 
with the disk placed at the center of a unit square.
We will discover that if we formulate $p(l)$ in terms of the rescaled variable
$\xi =2Rl$ then $p(l(\xi)$ will have a well defined limit 
(if we provide $p(\l)$ with some smearing)
when
$R$ gets small, and the following function
\begin{equation}
c(R)= \int \log \xi \cdot p(l(\xi))\; d\xi /2R    \label{eqn:cR}
\end{equation}
will have a well defined limit
and the Lyapunov exponent will be
\begin{equation}
\lambda=-2\log R -\log 2+1+c(R)+O(R)   \ \ .   \label{eqn:31}
\end{equation}

The whole problem thus lies in computing the function 
$p(l(\xi ))$ in the limit of small
$R$.

\subsection{Calculation of the distribution of recurrence times}

It is much more easy to grasp the
geometry of the billiard if it is represented as a regular Lorentz gas
on a square lattice. 
The midpoints of the disks are sitting on lattice points in
an integer lattice. To each lattice point is associated
a lattice vector ${\bf q}=(m,n)$ . 
An immediate observation is that only disks represented by coprime
lattice vectors ($gcd(m,n)=1$ where {\em gcd} means {\em greatest common
divisor}) may be
reached.
We send out a ray from the 
$(0,0)$ disk.
The phase space of the surface of section can now be partitioned into
subsets $\Omega_{\bf q}$;
a phase space point $x_s \in \Omega_{\bf q}$ if the trajectory starting at
$x_s$ hits disk ${\bf q}$. 

For finite disk radii not all coprime disks are accessible, some
$\Omega_{\bf q}=$\O. 
To deal with this
problem we will first construct a scheme for enumerating the
coprimes.

\subsubsection{Enumerating coprimes, or climbing the Farey tree}

Without any loss a generality we restrict the considerations 
to the lattice points
$(m,n)$ in the first octant, that is $m\geq n\geq 0$.
Because of the restriction $gcd(m,n)=1$ there is a one-to-one correspondence
between these coprime lattice vectors ${\bf q}=(m,n)$ 
and rational numbers $n/m$. 

The set of all coprime vectors ${\bf q}=(m,n)$ with
$m\leq M$ is called the {\em Farey sequence} of order $M$. 
Next define the cross-product between ${\bf q}_1=(m_1,n_1)$ and
${\bf q}_2=(m_2,n_2)$ as ${\bf q}_1  \times {\bf q}_2=m_1 n_2-n_1 m_2$.
Consider a vector ${\bf q}=(m,n)$ an draw a line from $(0,0)$ to it.
Call the lattice point closest to the line and above ${\bf q}'$, 
and below is ${\bf q}''$ the neighbors of ${\bf q}$
in the Farey sequence of order $m$.
The Farey theorem then states that ${\bf q} \times {\bf q}'=+1$
and (for symmetry reasons) ${\bf q} \times {\bf q}''=-1$.
Geometrically this means that the distance from ${\bf q}'$ and ${\bf q}''$
to the line is $1/|{\bf q}|$.
A consequence of Farey's theorem and the fact that ${\bf q}$ is coprime
is that ${\bf q}={\bf q}'+{\bf q}''$.
We will call ${\bf q}'$ and ${\bf q}''$ the mother and father
of ${\bf q}$.

The coprime lattice vectors, or the rationals,
may be organized in the Farey tree \cite{Ftree}, see fig 1a.
There is a link between two vectors ${\bf q}'$ and ${\bf q}''$
if and only if they are Farey neighbors, that is $|{\bf q}'\times {\bf q}''|=1$.
From each vector there are two links going upwards (towards the root as
the tree is drawn upside down), one up to the right to the mother
and one up to the left to the father. One of the parents lie on an adjacent
level, the other is not.

We now claim the following:

{\em Every coprime lattice vector ${\bf q}$, except $(1,0)$, $(2,1)$ and 
$(1,1)$, can uniquely
be written as ${\bf q}={\bf q}'+n{\bf q}_c$ 
where $n\geq 2$ and ${\bf q}'$ is the mother or father of
${\bf q}_c$}

Let's assume it is possible and show how to do the construction.
Call ${\bf q}\equiv {\bf q}_n={\bf q}'+n{\bf q}_c$.
By assumption we have $|{\bf q}'\times {\bf q}_c|=1$.
It follows that ${\bf q}_n$ is linked to ${\bf q}_c$ for all $n$.
It also follows that ${\bf q}_n$ is linked to ${\bf q}_{n-1}$ which
is linked to ${\bf q}_{n-2}$ etc.
The genealogy must thus be as in fig 1b.

So, given a vector ${\bf q}$ we do as follows to uniquely determine
$n$, ${\bf q}_c$ and ${\bf q}'$.
First we identify the parents of ${\bf q}$. One is on the level just above but
the other is not. The rule $n\neq 1$ force us to choose the other
as ${\bf q}_c$. Now we construct ${\bf q}_{n-1}$, ${\bf q}_{n-2}$
iteratively until we reach 
${\bf q}_{n-m}={\bf q}_0={\bf q}'$ for some $m$. 
It is easy to identify we this happen, since ${\bf q}_1$ is on a level
below ${\bf q}_c$ and ${\bf q}_0$ is located above it.
(Note that the dashed line in fig 2b is forbidden so this stopping criterion is
indeed unique).

\subsubsection{The size of $\Omega_{\bf q}$}

We want to compute the distribution \EqRef{pl}. We then split up the 
phase space integral
according to the partition into subsets $\Omega_{\bf q}$ 
\begin{equation}
p(l)=\sum_{\bf q} \int_{\Omega_{\bf q}} 
\delta (l-l_s(x_s)) dx_s  \ \ . \label{eqn:psoss}
\end{equation}
where the sum runs over all coprime lattice vectors.
We will now smear this distribution
\begin{equation}
p_\sigma(l)=\sum_{\bf q} \int_{\Omega_{\bf q}} 
\delta_\sigma (l-l_s(x_s)) dx_s  \ \ . \label{eqn:psoss}
\end{equation}
The smearing functions $\delta_\sigma(x)$ may be choosen as gaussians (this is
not essential, the important thing is that the flanks decay at least
exponentially). The mean and variance is given by
\begin{equation}\begin{array}{c}
\int x\delta_\sigma(x)dx=O(\sigma)\\
\int x^2\delta_\sigma(x)dx=O(\sigma^2)  \label{eqn:psm}
\end{array}  \ \ .
\end{equation}
The exact mean and variance is left undeterminate and may be
different from one place to another. This will enable us to write
equalities such as $\delta_\sigma(x+\epsilon)=\delta_\sigma(x)$
as long as $\epsilon$ does not exceed $\sigma$ in magnitude.

We will choose $\sigma \sim 1$. We will eventually consider $p_\sigma(l)$ as a
function of $\xi=2R l$. The smearing width in $\xi$-space is thus $\sim R$.
The goal is to average $\log \xi$ over 
$p(l(\xi))$. The error
induced by exchanging $p(l)$ with the smeared version $p_\sigma(l)$ will be
discussed in section 2.4. 

We will also assume some ambiguity
concerning the normalization of the smearing functions
\begin{equation}
\int \delta_\sigma(x)dx=1+O(R)  \label{eqn:normfuzz}  \ \ .
\end{equation}

We will now rewrite eq. \EqRef{psm} using our conventions on the smearing
function.
If $x_s\in \Omega_{\bf q}$ then $l_s(x_s)=q+O(R)$ and we write  
\begin{equation}
p_\sigma(l)=\sum_{\bf q} \int_{\Omega_q}  
\delta_\sigma (l-q+O(R)) =
 \sum_{\bf q} a_{\bf q}  
\delta_\sigma (l-q) \ \ . \label{eqn:p}
\end{equation}
where
\begin{equation}
a_{\bf q}=\int_{\Omega_q} dx_s
\end{equation}
and $q=|{\bf q}|$. Our goal here is to compute
$a_{\bf q}$.
  
Consider a trajectory hitting disk ${\bf q}$.
The relation between the phase space variables $\phi$ and $\alpha$
and the scattering angle $\beta_{\bf q}$ on disk ${\bf q}$ is (see fig 2). 
\begin{equation}
 \frac{q}{R}\sin (\phi - \theta_{\bf q}-\alpha)+\sin\alpha=\sin
\beta_{\bf q}  \ \ , \label{eqn:olikhet}
\end{equation}
where $\theta_{\bf q}$ is the polar angle of the lattice vector ${\bf q}$.
Provided that there are no interposed disks $\Omega_{\bf q}$ 
is given by $-1<\sin \beta_{\bf q} <1$ and $-1<\sin \alpha <1$.
Expanding
$\sin (\phi - \theta_{\bf q}-\alpha)$ we get 
\begin{equation}
 \frac{q}{R} (\phi - \theta_{\bf q}-\alpha) +O(R^2/q^2)+\sin\alpha=\sin
\beta_q  \ \ , \label{eqn:appolikhet}
\end{equation}

It is convenient to change integration variables in $dx_s=d\phi d(\sin
\alpha)$ from $(\phi ,\sin ( \alpha)) $ to
$(\sin ( \beta_q)) ,\sin ( \alpha))$. We thus get
\begin{equation}
dx_s=\frac{ R}{4\pi q}d( \sin \alpha)d( \sin  
\beta_{\bf q})(1+O(R^2/q^2))  \ \ . 
\end{equation}

\vspace{0.5cm}

First consider the case $q<1/2R$:\\
As we showed in ref. \cite{PDsin} disk ${\bf q}$ is then
fully accessible and
\begin{equation} \begin{array}{lll}
a_q=\frac{R}{\pi q}(1+O(R^2)) & \mbox{  ,  } & q<\frac{1}{2R}  \ \ .
\end{array}
\end{equation}

\vspace{0.5cm}
 
Next consider disks lying beyond the horizon $q>1/2R$:\\
We make use of the results of section 2.3.1 and write
${\bf q}_n={\bf q}'+n{\bf q}_c$.
We say that disk ${\bf q}_n$ lies in the ${\bf q}_c$ {\em corridor}.
Note that to each ${\bf q}_c$ the corresponds two ${\bf q}'$
one with smaller and one with larger polar angle (preceeding and succeeding
${\bf q}_c$
in the appropriate Farey sequence).
Below we assume that ${\bf q}'$ is the one with the larger polar angle, the
other case is completely analogous.
The geometry in the ${\bf q}_c$ corridor is illustrated in fig 3.
If $\Omega_{{\bf q}_n}$ were not obscured it would be given
by 
\begin{equation}
 \frac{q_n}{R} (\phi  -\alpha)
+O(R^4)+\sin \alpha =\sin  \beta_{{\bf q}_n}   \ \ \label{eqn:noll},
\end{equation}
with 
${q_n=|\bf q}'+n{\bf q}_c|$. Without loss of generality we can set
$\theta_{{\bf q}_n}=0$. The reason why
$O(R^2/q_n^2)$ in  \EqRef{appolikhet} becomes
$O(R^4)$ here is that $q_n>1/2R$ by assumption.
From now on we set $\beta_{{\bf q}_n}\equiv \beta$ as there is no
risk of confusion. 

The disk ${\bf q}_n$ under observation is shadowed only by two disks, namely
 ${\bf q}_{n-1}$
and ${\bf q}_c$.
The question is now where the phase space of these cut through the
$(\alpha,\beta)$-plane corresponding to ${\bf q}_n$. First we consider
${\bf q}_c$. The relevant border is described by an equation obtained
from \EqRef{appolikhet}.
\begin{equation}
 \frac{q_c}{R} (\phi +\gamma_n-\alpha)
+O(R^4)+\sin\alpha=1  \ \ , \label{eqn:tvaa}
\end{equation}
We now combine \EqRef{noll} and \EqRef{tvaa} to eliminate
$\phi$ and get
\begin{equation}
q_n(1+O(R^4))=q_c\sin\beta+(q_n-q_c)\sin\alpha+\frac{1}{R}q_c q_n \gamma_n  
\ \ . 
\end{equation}
With some elementary geometry one may show that
$q_c q_n \gamma_n=1+O(R^2)$ (to get the $O(R^2)$ correction one has to use that
$q_c\geq 1$ and $q_n\leq1/2R$),
so
\begin{equation}
q_n(1+O(R^4))=q_c\sin\beta +(q_n-q_c)\sin \alpha +\frac{1}{R}(1+O(R^2))
  \ \ .  \label{eqn:two}
\end{equation}
Next we do the same thing with ${\bf q}_{n-1}$. The equation for the borderline
is
\begin{equation}
\frac{q_{n-1}}{R}(\phi-(\gamma_{n-1}-\gamma_n)-\alpha)+\sin\alpha+O(R^4)=-1
  \ \ .  \label{eqn:ett}
\end{equation}
Combining  \EqRef{ett} and \EqRef{noll} we get
\begin{equation} 
-q_n(1+O(R^4))=q_{n-1}(\sin \beta -\sin \alpha)+q_n \sin \alpha-
\frac{1}{R} q_{n-1}q_n(\gamma_{n-1}-\gamma_n)  \ \ .   \label{eqn:one}
\end{equation}
Another geometry exercise shows that
$q_{n-1}=q_{n}-q+O(R^2)$ and
$q_{n-1}q_n(\gamma_{n-1}-\gamma_n)=1+O(R^4)$ so we arrive at
\begin{equation} 
-q_n(1+O(R^4)=(q_n-q_c)\sin \beta +q_c \sin \alpha-
\frac{1}{R}(1+O(R^4))  \ \ .
\end{equation}
To leading order in $R$  \EqRef{two} and \EqRef{one}
describe straight lines cutting through the $(\sin \beta \sin
\alpha)$-plane. 
Provided that  $q_c<1/2R$ there is a non vanishing area in upper-left corner
in the $(\sin \beta, \sin \alpha)$-plane for any $q_n$
(lower left corner if ${\bf q}'$ is chosen as the Farey predecessor). 
So we arrive at the following results 
\begin{equation}
a_q=\frac{R}{\pi q}(1+O(R^2))Ê\cdot \left\{ \begin{array}{ll}
1 & q<\frac{1}{2R} \\
(1-\frac{(1/2R-q)^2}{q_c(q-q_c)}) & 
\frac{1}{2R}<q<\frac{1}{2R}+q_c\\
\frac{(1/2R-q_c)^2}{(q-q_c)(q-2q_c)} & \frac{1}{2R}+q_c <q
\end{array} \right. \label{eqn:sumriz}
\end{equation}

\subsubsection{The density of coprimes}

The derivation below is very similar to the standard derivation of
the asymptotic behaviour of sums over
Eulers totient function $\sum_{n=1}^{N}\varphi(n)$, 
for details see e.g. \cite{Num}.

We are interested in the number of coprime lattice points inside
a certain radius $r$ (in this section we do not
restrict the consideration to the first
octant)
\begin{equation} 
N(r)=\sum_{\begin{array}{c} m,n\\m^2+n^2\leq r^2\\ gcd(m,n)=1\end{array}}1  
\ \ .
\end{equation}
We are going to relate this function the total number of lattice points inside
the circle of radius $r$
\begin{equation} 
R(r)=\sum_{\begin{array}{c} m,n\\m^2+n^2\leq r^2\end{array}}1  \ \ .
\end{equation}
This is possible since any vector $(m,n)$ may uniquely be written
as $(m,n)=d\cdot (\hat{m},\hat{n})$ where $d=gcd(m,n)$ and $(\hat{m},\hat{n})$ 
are coprime.
we can thus write
\begin{equation}
R(r)=\sum_{d=1}^{r} N(r/d)  \ \ ,
\end{equation}
Using Moebius inversion theorem we get
\begin{equation}
N(r)=\sum_{m=1}^{r}\mu(m) R(r/m)  \ \ ,
\end{equation}
where $\mu(m)$ is Moebius function.
The function $R(r)$ is
\begin{equation}
R(r)=\pi r^2+E(r)  \ \ ,
\end{equation}
where the error term,
according to a a theorem of Sierpinski, is $E(r)=O(r^{2/3})$ which means that
there exist a constant $C$ such that $|E(r)|<C\; r^{2/3}$ for all $r>1$.
Combining the two previous expression we get
\begin{equation}
N(r)=\pi r^2 \left( \sum_{m=1}^{\infty}\mu(m) \frac{1}{m^2}-
\sum_{m=r+1}^{\infty}\mu(m) \frac{1}{m^2}\right)
+\sum_{m=1}^{\infty}\mu(m) E(r/m)  \ \ ,
\end{equation}
where the first infinite sum is related to Riemann's zeta function 
$\sum_{m=1}^{\infty}\mu(m)/m^2=1/\zeta(2)=6/\pi^2$.
Now to the second term
\begin{equation}
|\sum_{m=r+1}^{\infty}\mu(m) \frac{1}{m^2}|<
\sum_{m=r+1}^{\infty}\frac{1}{m^2}=O(1/r)  \ \ ,
\end{equation}
where we used $|\mu(m)|\leq 1$.
The third sum may be given a similar bound
\begin{equation}
|\sum_{m=1}^{\infty}\mu(m) E(r/m)|<\sum_{m=1}^{\infty}| E(r/m)|<
C\; r^{2/3} \sum_{m=1}^{r} \frac{1}{m^{2/3}}=O(r)  \ \ .
\end{equation}
So we arrive at the result
\begin{equation}
N(r)=\frac{6}{\pi}r^2+O(r)  \ \ .
\end{equation}
We will use the derivative of this function, which is the mean density of
coprimes with respect to the radius
\begin{equation}
d_c(r)=\frac{dN(r)}{dr}=\frac{12r}{\pi}+O(1)  \ \ .
\end{equation}
One should of course be careful when integrating asymptotic series like this,
but the procedure above is legitimate for any smearing 
of the exact
$N(r)$.

\subsubsection{Assembling the pieces}

In this section we are going sum up the different contributions
$a_{\bf q}$ computed in section 2.3.2 by means of the means density of
coprimes computed in section 2.3.3.

First we consider the case $l<1/2R$.\\
Then $a_{\bf q}$ is a function of $q$ only: 
$a_{\bf q}=a(q)=R/(\pi q)\cdot(1+O(R^2))$,
cf. eq. \EqRef{sumriz}
\[
p_\sigma(l)=\sum_{\bf q} a_{\bf q}  
\delta_\sigma (l-q)  =
\sum_{\bf q} \frac{R}{\pi q}\cdot(1+O(R^2))  
\delta_\sigma (l-q)
\]
\begin{equation}
=\sum_{\bf q} \frac{R}{\pi q} 
\delta_\sigma (l-q)
=d_c(l) \frac{R}{\pi l}=
\frac{12R}{\pi^2}+R\cdot O(1/l)   \ \ .
\end{equation}
We appealed to eq \EqRef{normfuzz} to get rid of the $O(R^2)$ term in going from
the first to the second line above. 
We reformulate the last expression slightly
\begin{equation}
p_\sigma(l(\xi))=
\frac{12R}{\pi^2}+O(R^2)\cdot O(1/\xi) \;  \;  \; 2R<\xi<1  \ \ ,
\end{equation}
where $\xi =2R l$.

\vspace{0.5cm}

Next we consider the {\em transition region} $1/2R<l<1/R$.\\
We then use the result of section 2.3.2
to write
${\bf q}={\bf q}'+n{\bf q}_c$. According to eq. \EqRef{sumriz}
the amplitudes $a_{\bf q}$
depend on the  size of corridor $q_c=|{\bf q}_c|$ and the length of {\bf q}: 
$a_{\bf q}=a(q_c,q)$.
The length $q$ is $q=\Delta q_c+nq_c+O(R)$, where $\Delta$ is defined in fig 3,
it is a function of ${\bf q}_c$. We get
\begin{equation}
p_\sigma(l)=2\sum_{{\bf q}_c}\sum_{n=2}^{\infty}a(q_c,q)=
\delta_\sigma(l-q)
\end{equation}
\[
=2\sum_{{\bf q}_c}\sum_{n=2}^{\infty}a(q_c,q)\delta_\sigma(l-q_cn-a+O(R))
  \ \ ,  
\]
where we inserted the factor 2 to account for both parents
${\bf q}'$ and ${\bf q}''$ of ${\bf q}_c$.
The $O(R)$ in the argument gets swallowed by the
width of the delta function. Now we will turn the sum over ${\bf q}_c$ into an
integral over the density of coprime lattice vectors
\begin{equation}
p_\sigma(l)=2\int_0^{1/2R} dq_c\; d_c(q_c) \; a(q_c,l) \sum_{n=2}^{\infty}
\int_0^1
d\Delta f_{q_c}(\Delta)\delta_\sigma(l-q_cn-q_c\Delta) \ \ ,
\end{equation}
where $f_{q_c}(\Delta)$ is the distribution of the parameter $\Delta$.
Since this gets smeared by the width $\sigma/q_c\sim 1/q_c$ the results of
the Appendixcan be applied and $f_{q_c}(\Delta)$ can be replaced by a uniform
distribution $f_{q_c}(\Delta) \sim 1$
\[
p_\sigma(l)=2\int_0^{1/2R} dq_c\; d_c(q_c) \; a(q_c,l) \int_2^\infty
d\eta \delta_\sigma(l-q_c\eta)
\]
\begin{equation}
=2\int_0^{1/2R} dq_c\; d_c(q_c) \; a(q_c,l)\frac{1}{q_c} 
\theta_\sigma(l-2q_c)
\end{equation}
\[
=2\int_0^{min(1/2R,l/2)} dq_c\; d_c(q_c) \; a(q_c,l)\frac{1}{q_c}  \ \ ,
\]
where $\theta_\sigma(x)$ is a smeared step function. 
The restriction the the range $l<1/R$ implies that
$min(1/2R,l/2)=l/2$.
Next we insert the expression for $a(q_c,q)$ from eqs. \EqRef{sumriz}.
We then dispose of the $O(R^2)$ as before and for the moment suffice
with leading term of $d_c(q_c)$
\begin{equation}
p_\sigma(l)=2\int_{l-1/2R}^{l/2}dq_c \; \frac{12q_c}{\pi}\; \frac{1}{q_c}
\frac{R}{\pi l}(1-\frac{(1/2R-l)^2}{q_c(l-q_c)})+
\end{equation}
\[
2\int_{0}^{l-1/2R}dq_c\;\frac{12q_c}{\pi}\; \frac{1}{q_c}
\frac{R}{\pi l}\frac{(1/2R-q_c)^2}{(l-q_c)(l-2q_c)}+E(\xi)  \ \ .
\]
We change integration variable to $\eta =2Rq_c$ and use as before
$\xi =2R l$ and we have to solve the integral
\[
p_\sigma(l(\xi))=\frac{24R}{\pi^2} \left(
\int_{\xi-1}^{\xi/2}\frac{d\eta}{\xi}(1-\frac{(1-\xi)^2}{\eta(\xi-\eta)})+
\int_0^{\xi-1}\frac{d\eta}{\xi}\frac{(1-\eta)^2}{(\xi-\eta)(\xi-2\eta)}
\right)+E(\xi)=
\]
\begin{equation}
\frac{6R}{\pi^2 \xi^2}(2\xi+\xi(4-3\xi)\log(\xi)+
4(\xi-1)^2 \log(\xi-1)-(2-\xi)^2\log(2-\xi))+E(\xi)  \ \ .
\end{equation}

\vspace{0.5cm}

Now remains only the case $l>1/R$:\\
The calculation is completely analogous
to the previous case and we get
\[
p_\sigma(l(\xi))=2\int_{0}^{1/2R}dq_c \; d_c(q_c)\frac{1}{q_c}\;a(q_c,q=l)
\]
\begin{equation}
=2\int_{0}^{1/2R}dq_c \; \frac{12q_c}{\pi}\; \frac{1}{q_c}
\frac{R}{\pi l}\frac{(1/2R-q_c)^2}{(l-q_c)(l-2q_c)}+E(\xi)
\end{equation}
\[
=\frac{24R}{\pi^2}
\int_0^{1}\frac{d\eta}{\xi}\frac{(1-\eta)^2}{(\xi-\eta)(\xi-2\eta)}+E(\xi)
\]
\[
=\frac{6R}{\pi^2 \xi^2}(2\xi+\xi(4-3\xi)\log(\xi)+
4(\xi-1)^2 \log(\xi-1)-(2-\xi)^2\log(\xi-2))+E(\xi)  \ \ .
\]
The $1<\xi<2$ and $\xi>2$ results can be summarized by, 
\begin{equation}
p_\sigma(l(\xi))= \left\{ \begin{array}{ll}
\frac{12R}{\pi^2}+E(\xi) & \xi<1\\
\frac{6R}{\pi^2 \xi^2}(2\xi+\xi(4-3\xi)\log(\xi)+
4(\xi-1)^2 \log(\xi-1)-(2-\xi)^2\log|2-\xi|)+E(\xi)&
\xi>1 \end{array} \right.     \ \ ,
\end{equation}
where $\xi= 2Rl$.
It is instructive to perform a Laurent series expansion of $p_\sigma(l)$ for
$\xi>2$.
\begin{equation} 
p_\sigma(l(\xi))= \frac{48R}{\pi^2}\sum_{k=1}^{\infty}
\frac{2^k-1}{k(k+1)(k+2)}\frac{1}{\xi^{k+2}}    \ \ .  \label{eqn:Laurent}
\end{equation}

We now have an excellent opportunity to check our calculations
by computing the normalization factor
$\int p_\sigma(l) dl$. After a minor orgy of integration it is comforting
to discover that this factor is exactly unity.
The expectation value
$\int l p_\sigma(l) dl$ is found to be $1/2R$, 
which is correct up to $O(R)$.

Now remains only the error term $E(\xi)$. 
The remaining error is now exclusively
due the $O(1)$ term in $d_c(r)=12r/\pi+O(1)$. 
The case
$2R\leq \xi\leq 1$ is already clear.
The case $1\leq \xi \leq 2$ is also simple since there
$a_{\bf q}\leq R^2/2\pi$ so $E(\xi)=O(R^2) O(\xi^0)$.
The remaining case $\xi\geq 2$ needs a little more care since we would like to
keep the $1/\xi^3$ behaviour.
\begin{equation}
|E(\xi)|<CR^2\int_{2R}^{1} \frac{d\eta}{\eta}
\frac{(1-\eta)^2}{\xi(\xi-\eta)(\xi-2\eta)}\leq
\frac{2CR^2}{\xi^3}\int_{2R}^{1}\frac{1-\eta}{\eta}d\eta=O(R^2\log R) O(1/\xi^3)
  \ \ .
\end{equation}
We summarize
 
\begin{equation}
E(\xi)= \left\{ \begin{array}{ll}
O(R^2) O(1/\xi) & 2R<\xi<1\\
O(R^2) O(1)& 1<\xi<2 \\
O(R^2\log R) O(1/\xi^3) &2<\xi \end{array} \right.    \ \ .
\end{equation}

\subsection{The small $R$ limit of the Lyapunov exponent}

Now we will finally compute the integral
$\int \log \xi \cdot p(\xi)d\xi/2R$
without boring the reader with details.
When integrate over the region
$2\leq \xi$ we use the expansion \EqRef{Laurent} and express the integral
in terms of the sums
\begin{equation}\begin{array}{rl}
\sum_{k=1}^{\infty} \frac{1}{k^2}=&\zeta(2)=\pi^2/6\\
\sum_{k=1}^{\infty} \frac{1}{k^3}=&\zeta(3)=1.2020569031\ldots\\
\sum_{k=1}^{\infty} \frac{1}{2^k k}=&\log 2\\
\sum_{k=1}^{\infty} \frac{1}{2^k k^2}=&\pi^2/12-\frac{1}{2}\log^2 2\\
\sum_{k=1}^{\infty} \frac{1}{2^k k^3}=&
-\log 2 \frac{\pi^2}{12}+\frac{1}{6}\log^3 2+\frac{7}{8}\zeta(3)
\end{array}  \ \ .
\end{equation}
After this
we might as well
express the rest of the integral in terms of these sums.
The resulting $c(R)$ (as defined in eq \EqRef{cR}) is 
\begin{equation}
c(R)=
\int \log \xi p(\xi)d\xi/2R=\int \log \xi p_\sigma(\xi)d\xi/2R+E_1(R)
=\frac{27}{2\pi^2}\zeta(3)-3\log 2+E_1(R)+E_2(R)  \ \ ,
\end{equation}
where $E_1(R)$ is the error induced by smearing the delta functions, cf section
2.3.2, this may be shown to be $E_1(R)=O(R \log R)$.
The second error term is 
$E_2(R)=\int_{2R}^{\infty}\log \xi E(\xi)d\xi/2R=O(R\log^2 R)$  
where the dominating contribution comes from the
vicinity of  $\xi \sim 2R$.

We can now state our main result, cf. eq. \EqRef{31}, which is
\begin{equation}
\lambda=-2\log R+C+O(R\log^2R) \label{eqn:C}  \ \ ,
\end{equation}
with
\begin{equation}
C=1-4\log 2+\frac{27}{2\pi^2}\zeta(3)\approx-0.12837205  \ \ .
\end{equation}

We stress that the derivation in the appendix is still not rigorous and
the error estimate $O(R \log^2 R)$ is only conjectural.

\section{Discussion}

We stress that the approach the considerations in section 2 applies to
a large class of system but the function $p(l)$ need to worked out
for each particular geometry.
The term $-2\log R$ appears to be universal for two-dimensional
Sinai-billiards \cite{Oono,Chern,Dorf}.
One may use rather crude methods \cite{PDlyap}
to estimate $p(l)$ and get
quite decent estimates of the constant $C$ in eq. \EqRef{C}.

The distribution of recurrence times contains a lot of dynamical information,
concerning e.g. 
correlation decay and diffusion behaviour \cite{PDsin,DAcorr,PDlyap}.

I would like to thank Predrag Cvitanovi\'{c} for comments on the manuscript.
This work was supported by the Swedish Natural Science
Research Council (NFR) under contract no. F-FU 06420-303.

\newcommand{\PR}[1]{{Phys.\ Rep.}\/ {\bf #1}}
\newcommand{\PRL}[1]{{Phys.\ Rev.\ Lett.}\/ {\bf #1}}
\newcommand{\PRA}[1]{{Phys.\ Rev.\ A}\/ {\bf #1}}
\newcommand{\PRD}[1]{{Phys.\ Rev.\ D}\/ {\bf #1}}
\newcommand{\PRE}[1]{{Phys.\ Rev.\ E}\/ {\bf #1}}
\newcommand{\JPA}[1]{{J.\ Phys.\ A}\/ {\bf #1}}
\newcommand{\JPB}[1]{{J.\ Phys.\ B}\/ {\bf #1}}
\newcommand{\JCP}[1]{{J.\ Chem.\ Phys.}\/ {\bf #1}}
\newcommand{\JPC}[1]{{J.\ Phys.\ Chem.}\/ {\bf #1}}
\newcommand{\JMP}[1]{{J.\ Math.\ Phys.}\/ {\bf #1}}
\newcommand{\JSP}[1]{{J.\ Stat.\ Phys.}\/ {\bf #1}}
\newcommand{\AP}[1]{{Ann.\ Phys.}\/ {\bf #1}}
\newcommand{\PLB}[1]{{Phys.\ Lett.\ B}\/ {\bf #1}}
\newcommand{\PLA}[1]{{Phys.\ Lett.\ A}\/ {\bf #1}}
\newcommand{\PD}[1]{{Physica D}\/ {\bf #1}}
\newcommand{\NPB}[1]{{Nucl.\ Phys.\ B}\/ {\bf #1}}
\newcommand{\INCB}[1]{{Il Nuov.\ Cim.\ B}\/ {\bf #1}}
\newcommand{\JETP}[1]{{Sov.\ Phys.\ JETP}\/ {\bf #1}}
\newcommand{\JETPL}[1]{{JETP Lett.\ }\/ {\bf #1}}
\newcommand{\RMS}[1]{{Russ.\ Math.\ Surv.}\/ {\bf #1}}
\newcommand{\USSR}[1]{{Math.\ USSR.\ Sb.}\/ {\bf #1}}
\newcommand{\PST}[1]{{Phys.\ Scripta T}\/ {\bf #1}}
\newcommand{\CM}[1]{{Cont.\ Math.}\/ {\bf #1}}
\newcommand{\JMPA}[1]{{J.\ Math.\ Pure Appl.}\/ {\bf #1}}
\newcommand{\CMP}[1]{{Comm.\ Math.\ Phys.}\/ {\bf #1}}
\newcommand{\PRS}[1]{{Proc.\ R.\ Soc. Lond.\ A}\/ {\bf #1}}
%


\newpage

\section*{Appendix}

In this appendix we will present evidence that the parameter $\Delta$,
as defined in fig 3,
is uniformly distributed between zero and unity.

Let ${\bf q}'({\bf q})$ be the mother (cf. section 2.3.1) of ${\bf q}$
and ${\bf q}'({\bf q})$ its father.
Further, let $\Delta'({\bf q})$  be the $\Delta$-variable 
(see fig. 3)
associated
with the mother and similarly with the father.
Let now $F_r(\Delta)$ be the (probability) distribution
of all $\Delta$'s associated with all coprime lattice points ${\bf q}$
such that $q=|{\bf q}|<r$
\begin{equation}
N(r)\cdot F_r(\Delta)=
\sum_{\bf q} \theta(r-q)\frac{1}{2}\left(
\delta_\omega (\Delta-\Delta'({\bf q}))+
\delta_\omega (\Delta-\Delta''({\bf q})\right) 
\end{equation}
where $\omega$ is a smearing width to be defined later.
As in section 2.3.3 we also assume smearing of the step function we do not
specify it.
Let us further define
\begin{equation}
d(r)\cdot f_r(\Delta)=\frac{d}{dr} N(r)\cdot F_r(\Delta)
\end{equation}
where $F_r(\Delta)$ and $f_r(\Delta)$ will share their leading asymptotic
behaviour.

Next we use the result of section 2.3.1 to write each ${\bf q}$
as either ${\bf q}={\bf q}'({\bf q_c})+n{\bf q}_c$ or
${\bf q}={\bf q}''({\bf q_c})+n{\bf q}_c$.
Let us consider the entire sequence
${\bf q}_n={\bf q}'({\bf q_c})+n{\bf q}_c$.
Then ${\bf q}'({\bf q_n})={\bf q}_{n-1}$ and
${\bf q}''({\bf q_n})={\bf q}_c$ and
\[
\Delta'({\bf q}_n)=\frac{|{\bf q}_{n-1}|}{|{\bf q}_{n}|}+
O(1/q_{n}^2)
=
\frac{\Delta'({\bf q}_c)+n-1}{\Delta'({\bf q}_c)+n}+O(1/q_{n}^{2})
\]
\begin{equation}
\Delta''({\bf q}_n)=\frac{|{\bf q}_c|}{|{\bf q}_{n}|}+O(1/q_{n}^{3})=
\frac{1}{\Delta'({\bf q}_c)+n}+O(1/q_{n}^{3})
\end{equation}
\[
q_n=(\Delta'({\bf q}_c)+n)q_c+O(1/(nq_c^3)
\]
And we can write
\begin{equation}
N(r)\cdot F_r(\Delta)=\sum_{{\bf q}_c} \sum_{n=2}^{\infty}
\theta(r-(\Delta'({\bf q}_c)+n)q_c+O(1/(nq_c^3)))\cdot
\end{equation}
\[
\left\{\delta_\omega(\Delta-
\frac{\Delta'({\bf q}_c)+n-1}{\Delta'({\bf q}_c)+n}+O(1/(nq_c)^2))+
\delta_\omega(\Delta-
\frac{1}{\Delta'({\bf q}_c)+n}+O(1/(nq_c)^2)) \right\} 
\]
where we have used the fact that the sequence
${\bf q}={\bf q}''({\bf q_c})+n{\bf q}_c$ give rise to an equivalent term.

Next we insert the identities
\begin{eqnarray}
1=\int\delta(r'-q_c)dr'\\
1=\int_0^1 \delta_\omega(\Delta'-\Delta'({\bf q}_c))d\Delta'
\end{eqnarray}
After some work we arrive at
\begin{equation}
N(r)\cdot F_r(\Delta)=\int dr' \int_2^\infty dy
\theta(r-yr') d(r')f_{r'}(y \mbox{ mod } 1) \cdot \label{eqn:second}
\end{equation}
\[
\left\{ \delta_\omega(\Delta-\frac{y-1}{y}+O(1/r'^2))+
\delta_\omega(\Delta-\frac{1}{y}+O(1/r'^2)) \right\} 
\]
We now try to solve this integral equation asymptotically.
We may restrict ourselves to the case $\Delta<1/2$ since $P_r(\Delta)$
is symmetric around $\Delta=1/2$. 
We then only need to consider the second delta-function in \EqRef{second}.
We consider now only the contribution from lattice points
$r_1<q<r_2$ where $r_1\sim r_2\sim r$ is of the same order of magnitude.
To get rid of the error term in the argument of the delta function
we need to take $\omega =O(1/r^2)$
\begin{equation}
N(r_2)\cdot F_{r_2}(\Delta)-N(r_1)\cdot F_{r_1}(\Delta)
\end{equation}
\[
=\int dr' \int_2^\infty dy
\left( \theta(r_2-yr')-\theta(r_1-yr')\right) d(r')f_{r'}(y \mbox{ mod } 1)
\delta_\omega(\Delta-\frac{1}{y}) 
\]

We insert the leading asymptotic solutions 
$N\sim 6r^2/\pi$ and
$d \sim 12r/\pi$ and seek the leading distribution
$F_r(\Delta)\sim f_r(\Delta) \sim \rho(\Delta)$.
The integral equation then gives
\begin{equation}
\left( \frac{6r_2^2}{\pi}-\frac{6r_1^2}{\pi}\right)
\rho(\Delta)
\end{equation}
\[
=
\left( \frac{6r_2^2}{\pi}-\frac{6r_1^2}{\pi}\right)
\int_0^{1/2}dx  \delta_\omega(\Delta-x)\rho(1/x \mbox{ mod } 1)
\]
which is a smeared version of the functional equation
\begin{equation}
\rho(\Delta)=\rho(1/\Delta \mbox{ mod } 1)
\end{equation}
which combined with the symmetry constarint
\begin{equation}
\rho(1-\Delta)=\rho(\Delta)
\end{equation}
has the obvious solution $\rho(\Delta)=1$.

What we have presented is of course not a complete proof but we have presented
evidence that the distribution of the variable $\Delta$ obtained from
all lattice points insider radius $r$ tends uniformly to a uniform
distribution when $r \rightarrow \infty$
provided that the smearing width $\omega=O( 1/r^2)$.

\newpage

\section*{Figure captions}

\FigCap{1}{a) The Farey tree. b)Part of the Farey three corresponding
to the sequence ${\bf q}_n={\bf q}'+n{\bf q}_c$.}

\FigCap{2}{Definition of phase space variables use in section
2.3.2.}

\FigCap{3}{The geometry of the ${\bf q}_c$ corridor.}

\end{document}